\documentclass[journal]{IEEEtran}
\IEEEoverridecommandlockouts
\usepackage{amsmath,amssymb,amsfonts}
\usepackage{algorithmic}
\usepackage{graphicx}
\usepackage{textcomp}
\usepackage{xcolor}
\usepackage{float}
\usepackage{array}
\usepackage{booktabs}
\usepackage{subfigure}
\usepackage{amsmath}
\usepackage{multirow}
\usepackage[caption=false,font=normalsize,labelfont=sf,textfont=sf]{subfig}

\usepackage{algorithm}
\usepackage{algorithmic}

\def\BibTeX{{\rm B\kern-.05em{\sc i\kern-.025em b}\kern-.08em
    T\kern-.1667em\lower.7ex\hbox{E}\kern-.125emX}}
\usepackage{cite}
\begin{document}
\bstctlcite{IEEEexample:BSTcontrol}
\title{A 14$uJ$/Decision Keyword Spotting Accelerator with In-SRAM-Computing and On Chip Learning for Customization\\
}
\author{\IEEEauthorblockN{Yu-Hsiang Chiang, Tian-Sheuan Chang, \textit{Senior Member, IEEE}, and  Shyh Jye Jou, \textit{Senior Member, IEEE}}
\thanks{Manuscript received February 7, 2022; revised April 5, 2022; accepted May 2, 2022.}
\thanks{This work was supported in part by Taiwan Semiconductor Manufacturing Company (TSMGrant  and in part by the Ministry of Science and Technology, Taiwan, under Grant 110-2634-F-009-017,Grant 109-2639-E-009-001, Grant 110-2221-E-A49-148-MY3, and Grant 110-26Grant 22-8-009-018-SB. The authors are with the Institute of Electronics, National Yang Ming Chiao Tung University, Hsinchu 30010, Taiwan (e-mail: q123783293@gmail.com, tschang@nycu.edu.tw, jerryjou@g2.nctu.edu.tw) }%
\thanks{
© 2022 IEEE.  Personal use of this material is permitted.  Permission from IEEE must be obtained for all other uses, in any current or future media, including reprinting/republishing this material for advertising or promotional purposes, creating new collective works, for resale or redistribution to servers or lists, or reuse of any copyrighted component of this work in other works.\\
Y. -H. Chiang, T. -S. Chang and S. J. Jou, "A 14$uJ$/Decision Keyword Spotting Accelerator with In-SRAM-Computing and On Chip Learning for Customization," in IEEE Transactions on VLSI systems, 2022. doi: 10.1109/TVLSI.2022.3172685.
}
}
\maketitle

\begin{abstract}
Keyword spotting has gained popularity as a natural way to interact with consumer devices in recent years. However, because of its always-on nature and the variety of speech, it necessitates a low-power design as well as user customization. This paper describes a low-power, energy-efficient keyword spotting accelerator with SRAM based in-memory computing (IMC) and on-chip learning for user customization. However, IMC is constrained by macro size, limited precision, and non-ideal effects. To address the issues mentioned above, this paper proposes bias compensation and fine-tuning using an IMC-aware model design. Furthermore, because learning with low-precision edge devices results in zero error and gradient values due to quantization, this paper proposes error scaling and small gradient accumulation to achieve the same accuracy as ideal model training. The simulation results show that with user customization, we can recover the accuracy loss from 51.08\% to 89.76\% with compensation and fine-tuning and further improve to 96.71\% with customization. The chip implementation can successfully run the model with only 14$uJ$ per decision. When compared to the state-of-the-art works, the presented design has higher energy efficiency with additional on-chip model customization capabilities for higher accuracy.

\end{abstract}


\maketitle

\section{Introduction}

Motivated by the breakthrough of deep learning in speech recognition, voice recognition using keyword spotting (KWS) is a natural and increasingly popular way to interact with consumer devices. Since KWS is always on, it should have very low power for edge devices. 

Various works have been proposed for low-power KWS. Zhang et al.\cite{zhang2017hello} implements several different KWS models on micro controllers to compare their accuracy and memory/compute requirements. Zheng et al.\cite{zheng2019ultra} proposes a binary neural network-based design with on-chip self-learning to update the entire model. Dbouk et al.\cite{dbouk20200} uses recurrent attention network and hybrid digital and multibit in-memory computing for ultra low power KWS. Guo et al.\cite{guo20195} proposes hybrid digital circuits and 16 64x64 SRAM-based in-memory computing (IMC) macros with 3-bit ADC for recurrent neural network-based KWS. Liu et al.\cite{liu2019eera} uses precision self-adaptive computing for a binary weight network to reduce power, and \cite{liu2019ultra} uses mixed mode computing for low-power KWS. In summary, they reduce power consumption through recurrent models, quantized/binary neural network models, or voice activity detection. However, their model design did not consider the underlying hardware constraints or non-ideal effects for IMC. 

In addition to the low-power requirement, KWS models often face accuracy degradation due to the accent and pronunciation of different users in different regions. To recover accuracy, model personalization or customization is a popular technique for applications with data that vary significantly from person to person, such as KWS~\cite{warden2018speech}, human activity recognition~\cite{lin2020model}, and handwriting recognition~\cite{harris2018architectures}, which are demanded for edge AI devices. Model customization can be executed on either the chip or server side. For KWS, on-chip model customization is preferred over the server-side one due to the privacy concern to retrain the entire model or fine-tune a pre-trained model with a small amount of local data. However, edge devices usually use low-precision fixed-point hardware that poses a big problem for high-precision training. This situation is getting worse for fine-tuning a pre-trained model since the errors and gradients are quite small in such a case, and their quantization will lead to zero error and gradient. Thus, fine-tuning will lead to catastrophic failure.

To solve above issues, this paper proposes a low-power KWS chip with SRAM-based IMC and on-chip learning for customization. The model customization issue on low-precision hardware is solved by the proposed error scaling, 
small gradient accumulation, and random gradients, which can restore the accuracy as the full precision fine-tuning. IMC is adopted for its highly parallel computation and ultra-low power consumption. However, IMC also faces limited precision in the weight and activation and non-linearity effects of analog circuits. Thus, this paper proposes an IMC-aware model that uses a binary neural network with in-memory-batch normalization (BN) to minimize the conversion between digital and analog. Only the first and final layer uses digital implementation due to its higher precision needs. The non-ideal effects are solved by bias compensation and fine-tuning. The chip implemented shows higher energy efficiency compared to other state-of-the-art works with customization capability.

The remainder of the paper is organized as follows. Section II shows the baseline model for KWS. Section III presents the proposed on-chip training for model customization. Section IV shows the non-ideal effects of the IMC macro and how to solve them. Section V presents the proposed KWS chip architecture. Section VI shows the experimental results and comparisons. Finally, this paper is concluded in Section VII.

\section{Baseline Model for IMC aware KWS}

\begin{figure}[!htb]
  \centering
 \includegraphics[height=!,width=1.0\linewidth,keepaspectratio=true]
  {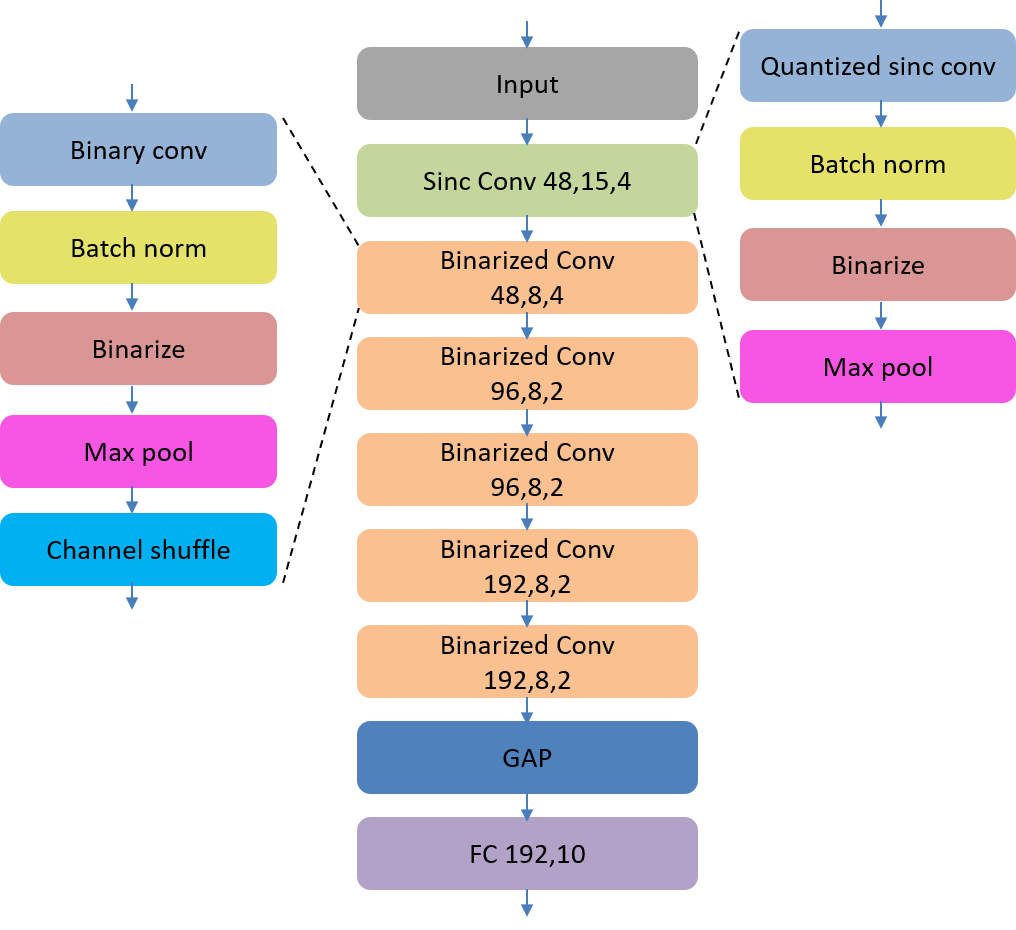}
  \caption[Model architecture of KWS.]{{The proposed IMC aware KWS model. The number in the block means output channel number, kernel size, pooling size from left to right.}}
  \label{fig:SRAM_model}
\end{figure}

Fig.~\ref{fig:SRAM_model} shows the overall IMC aware binary neural network model for KWS. The binary neural network is selected because the multibit ADC required for IMC could be simplified as a low area cost and low power sense amplifier (SA) when combined with in-memory BN~\cite{yonekawa2017chip}. The input is 8-bit raw audio data of nearly one second length. The model outputs the class of the input keyword. The model consists of one binarized sinc convolution layer \cite{ravanelli2018speaker}, which is a learned filter bank to process raw audio. Compared to conventional MFCC, the computational complexity will be lower, and this also makes the model an end-to-end learning model. Following the filter banks, five binary convolution layers are adopted that use group convolution with group size set to 24 and in-memory BN as the basic block. With in-memory BN, convolution and BN can be executed together within the array, and the array output can be a binary activation output that can be implemented by SAs instead of multibit ADCs. The weights of the sinc convolution layers are also binary for hardware consideration. Thus, the model uses only binary computation in the convolution layer in inference, while the final classifier layer uses 8-bit fixed-point computation. \par

\begin{figure}[!htb]
  \centering{\includegraphics[height=!,width=1.0\linewidth,keepaspectratio=true]{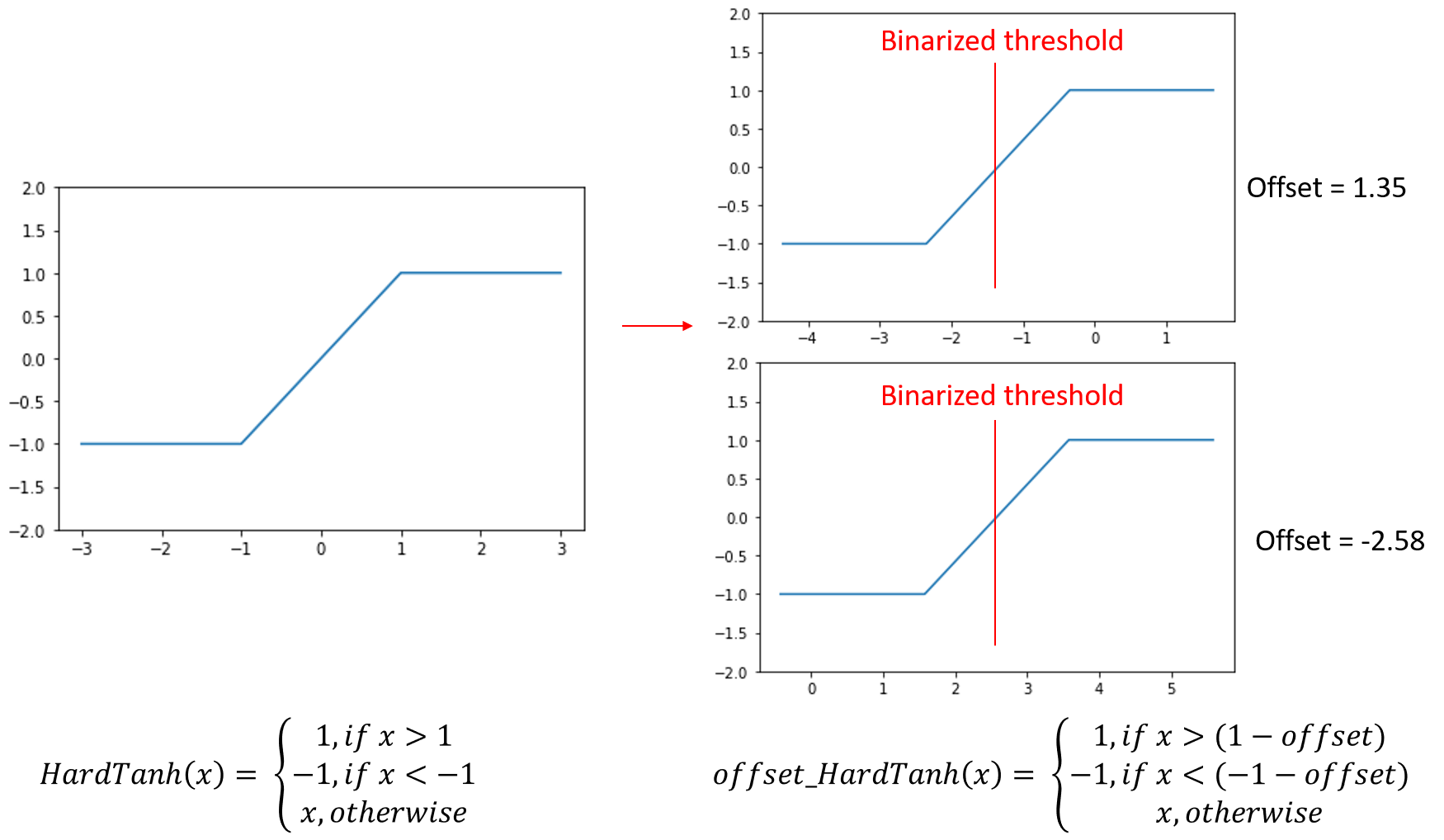}}
  \caption{A learnable offset before activation function to change the binarized threshold of different layer. For a positive offset(top right), more feature will be 1 after activation, for a negative offset(bottom right), more feature will be 0 after activation.}
  \label{fig:binarized_threshold}
\end{figure}

\begin{figure}[!htb]
  \centering
  \includegraphics[height=!,width=0.6\linewidth,keepaspectratio=true]
  {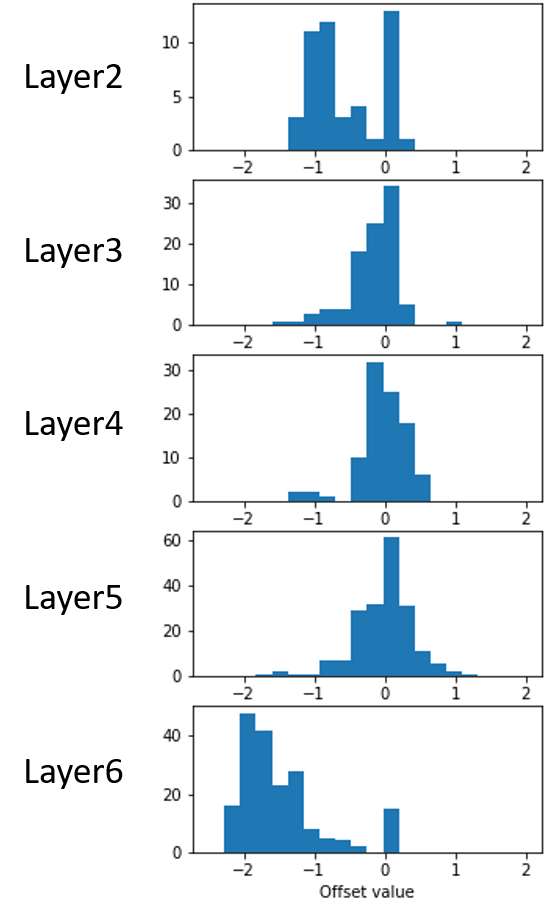}
  \caption{{Trained offset of each layer with initial zero value.}}
  \label{fig:offset_distribution}
\end{figure}

In addition to the above model, for better model training results, we adopt the trainable offset for binarized activation \cite{liu2020reactnet}, as shown in Fig.~\ref{fig:binarized_threshold}. This offset value can be merged with BN in the inference phase, which will not incur additional overhead for hardware implementation. Fig.~\ref{fig:offset_distribution} shows the trained offset value for each layer. Initial values are set to 0 for all layers. The figure shows that the appropriate offset is not the same for each layer, and the trainable offset can effectively preserve the extracted features. 

\section{On-chip training methods}

\subsection{Related Work}
For on-chip training and inference, the most recent studies show that training needs at least 8 bits of precision to ensure accuracy\cite{banner2018scalable, yang2020training, zhu2020towards} although precision can be extremely low at the inference phase. Most of the related work focuses on training models from scratch and building a software framework \cite{jacob2018quantization} for quantized training to reduce computational resources. Yang et al.\cite{yang2020training} observes the requirement of a large bit width to realize model convergence, and thus introduce a layer-wise scaling factor and an extra flag bit to solve the problem. Instead, this paper uses a more straightforward scaling method, as mentioned later. Zhu et al.\cite{zhu2020towards} suggests clipping the gradient value for integer training, but the clipping value is based on the cosine distance between the floating point gradient and its quantize-dequantized counterpart, which is impractical in the edge device.
Banner et al.\cite{banner2018scalable} proposes the Range BN for a higher tolerance to quantization noise, but collecting BN statistics has a high overhead on hardware. Furthermore, since a small training data set has too much variation, changing the BN statistics may make training unstable. Therefore, this paper decides to freeze the BN statistics during training.

\subsection{Model Customization}

For model customization, this paper fine-tunes the last classification layer, which can avoid computational overhead and storage for backpropagation of the internal layers. Additionally, the required feature map will be used right after the error calculation without storing it in a buffer. This will also be hardware-friendly. 

To implement this fine-tuning in hardware, we choose 8 bits fixed-point for the training part since fixed-point numbers are more hardware-friendly than floating-point numbers. 
However, when this on-chip fine-tuning is implemented with hardware-friendly quantized computation units, it is easy for the on-chip fine-tuning to fail because of the quantized value. In the following, we will propose hardware-oriented techniques to make on-chip fine-tuning more robust and hardware-friendly.\par

\subsection{Error scaling}
\begin{figure}[!htb]
  \centering {\includegraphics[height=!,width=0.8\linewidth,keepaspectratio=true]
  {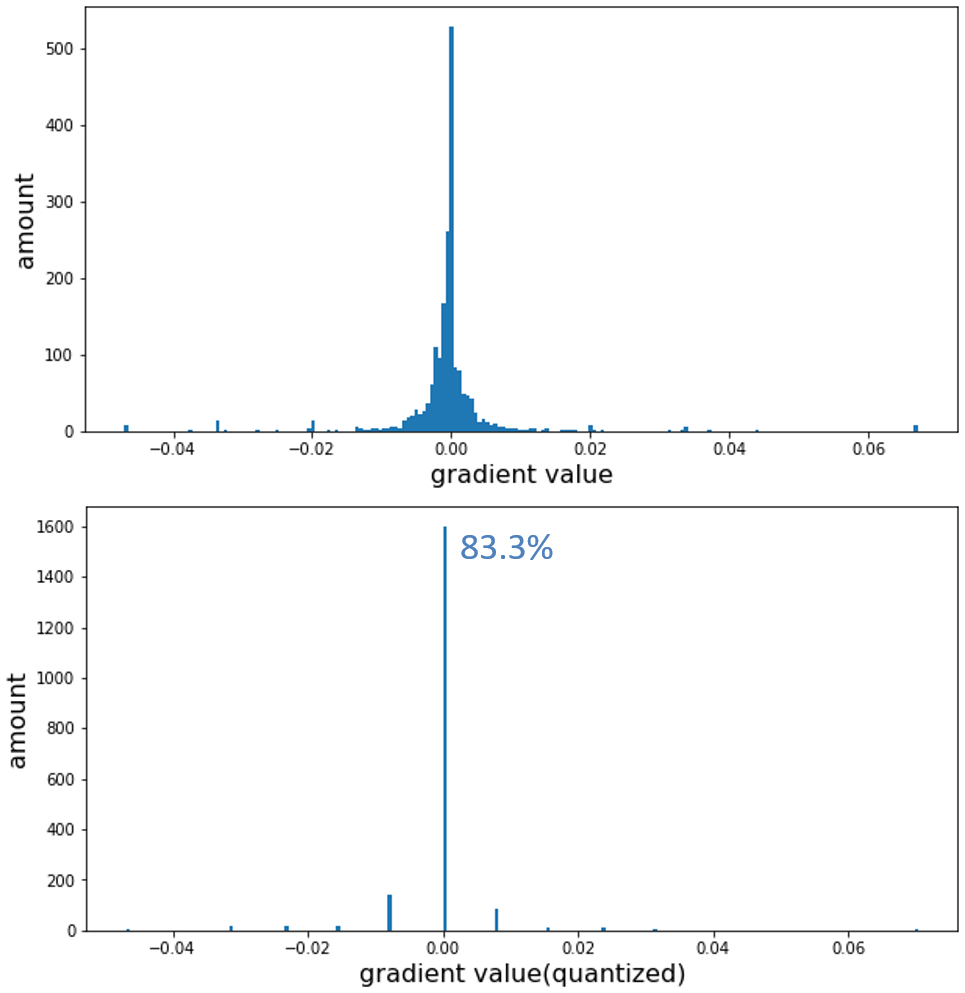}}
  \caption[Gradient distribution on the personal dataset.]{{ The Gradient distribution of our KWS model on the personal dataset before quantize(top) and after quantize(bottom).}}
  \label{fig:gradient_distribution}
\end{figure}

Training from scratch in a low-precision format has been proven to work well. However, when fine-tuning a pre-trained model, most of the error values will be close to zero, as shown in Fig.~\ref{fig:gradient_distribution}, since the model has converged well.  These small error values will be quantized to zero after quantization, causing the model not to learn any information from the personal data. Thus, we add a scaling factor before the error quantization. For a desired scale error \textit{ScaleError} as shown in (\ref{eq:scaling_error}), its scaling factor can be derived as (\ref{eq:scaling_factor}) to make training more general. 
This method does not need an extra flag bit, as in \cite{yang2020training} to indicate whether the absolute value is smaller than the scaling factor.

The error scaling works as follows. Assuming that we want the error values distributed between 1 and -1 to match the bit format, we need different scaling factors for different error distributions to make training more general. Therefore, we use (\ref{eq:scaling_error}) to scale the error values, where the scaling factor s is calculated by the extreme error value shown in (\ref{eq:scaling_factor}) to ensure that the distribution could be expressed completely by the limited bit format. This method does not need an extra flag bit, as in \cite{yang2020training} to indicate whether the absolute value is smaller than the scaling factor.

\begin{equation}
ScaleError = error * 2^{s}
\label{eq:scaling_error}
\end{equation}

\begin{equation}
s = \lceil log_2(\frac{1}{max(|error|)})\rceil
\label{eq:scaling_factor}
\end{equation}

\subsection{Small gradient accumulation}
As mentioned above, most of the gradient values are close to zero after quantization, and therefore the gradient values will be too small to change the model weights. In addition, these gradient values will become smaller updated weight values since the learning rate during fine-tuning is also quite small, which will make the model easily stop learning at the early training stage. 

To avoid this, we accumulate the gradient whose value is smaller than the threshold. Once the accumulated gradient is larger than a threshold, we use the accumulated gradient to update the weight and then reset the accumulated value to 0. These accumulated values are in 16 bit fixed-point format to ensure that the training will not use any full precision number. The pseudocode is shown in Algorithm ~\ref{alg:SGA}. In which the quantization threshold $G_{th}$ depends on the learning rate \textit{LR}. Eq.~(\ref{eq:gth}) shows the relationship, where \textit{min(weight)} means the minimum weight value that can be expressed. Table~\ref{table:G_th} shows some examples of the threshold value when the weight value is quantized to 1 sign bit and 7 decimal bits, which means that \textit{min(weight)} is 1/128.

\begin{algorithm}[h]
\caption{Small Gradient Accumulation(SGA)}
\begin{algorithmic}
\STATE $G: \mbox{The weights gradient value of this iteration}$
\STATE $G_{accu}: \mbox{The accumulated gradient value}$
\STATE $G_{th}: \mbox{The accumulation threshold}$
\STATE $G_{update}: \mbox{The gradient value used to update weights}$
\IF{$G < G_{th}$}
\IF{$G_{accu} < G_{th}$}
\STATE $G_{accu} \leftarrow G_{accu} + G$
\ELSE
\STATE $G_{update} \leftarrow G_{accu} + G$
\STATE $G_{accu} \leftarrow 0$
\ENDIF
\ELSE
\STATE $G_{update} \leftarrow G$
\ENDIF
\end{algorithmic}
\label{alg:SGA}
\end{algorithm}

\begin{equation}
G_{th} = \frac{min(weight)/2}{LR}
\label{eq:gth}
\end{equation}

\begin{table}[]
\centering
\caption[Examples for the threshold value with min(weight)=1/128.]{Examples for threshold values with min(weight)=1/128}
\label{table:G_th}
\begin{tabular}{@{}cccc@{}}
\toprule
          & LR=0.05                   & LR=0.01                   & LR=0.001                 \\ \midrule
Threshold & 0.078                     & 0.039                     & 0.39                     \\ \bottomrule
\end{tabular}
\end{table}

\subsection{Random gradient prediction(RGP)}
When the training data set is small enough, we can read all data in a single batch, which means that the input data for the last layer will be very close in each epoch. Thus, we add Gaussian noise to predict the gradient of the next epoch as (\ref{eq:RGP}).

\begin{equation}
G'= G + quantize(\frac{rand}{\lambda})
\label{eq:RGP}
\end{equation}

\begin{figure}[!!htb]
  \centering
  \includegraphics[height=!,width=0.8\linewidth,keepaspectratio=true]
  {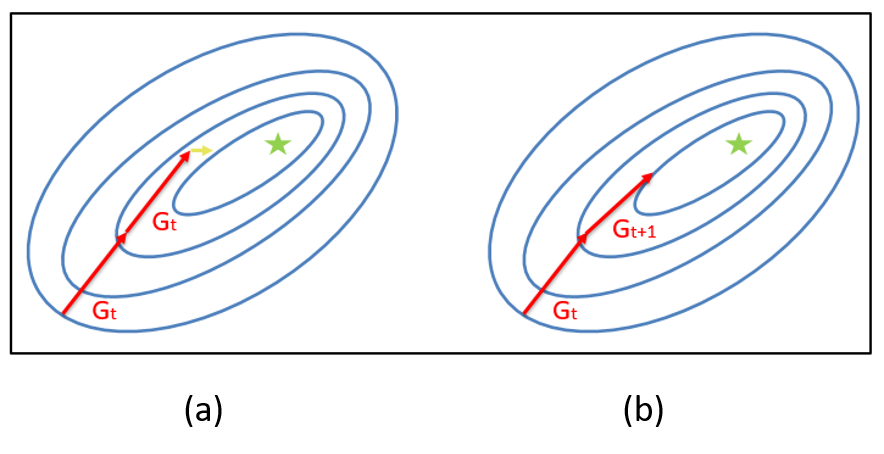}
  \caption[Random Gradient prediction.]{{ (a) Weight update with random gradient prediction, where the yellow arrow means the random direction. (b) Weight update without random gradient prediction.}}
  \label{fig:random_gradient_prediction}
\end{figure}

In (\ref{eq:RGP}), \textit{rand} is a random sample of the Gaussian distribution, and the value of $\lambda$ is a hyperparameter. With a suitable value of $\lambda$ , the noise value will not dominate the update direction and can avoid the model stuck at the local minimum. Another advantage is that we can ensure that the small truncated error caused by the hardware calculation will not affect the overall training of the model. Fig.~\ref{fig:random_gradient_prediction} illustrates this weight update method.

\section{Design for Non-ideal Effects of the IMC Macro}

\begin{figure}[htb]
  \centering
  \includegraphics[height=!,width=0.6\linewidth,keepaspectratio=true]
  {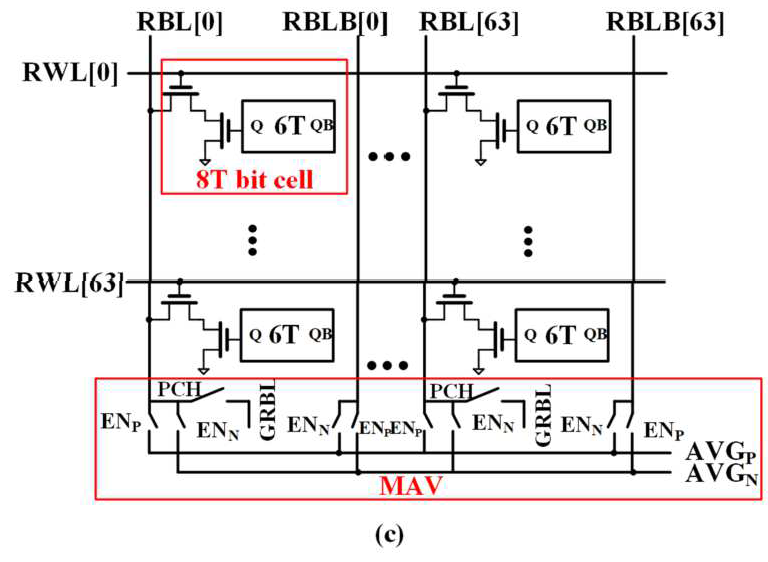}
  \caption{Architecture of the adopted IMC design\cite{jouisc2021}.}
  \label{fig:isc}
\end{figure}

The IMC macro used in this paper is based on the binary neural network macro from our previous work\cite{jouisc2021}. One IMC macro contains 8 $64\times64$ banks as Fig.~\ref{fig:isc} to compute eight outputs, which is equivalent to the size of 4KBytes. This macro uses 8T SRAM bit cells from the foundry to avoid a read disturb problem. For multiplication and accumulation in a convolution, weights are first written into the 8T SRAM array. The read bitlines (RBL) are precharged according to the input data. Then a wordline of weights are read from the SRAM to decide whether to discharge or keep the RBL voltage according to the weight as the multiplication results. These results are accumulated and averaged by charge sharing on AVG lines ($AVG_P$ and $AVG_N$) based on their sign. Finally, the AVG lines will be sent to SAs to convert them into 1-bit output results. For more details, see \cite{jouisc2021}. Due to this analog computing for multiply and average (MAV), IMC has some model design limitations and non-ideal effects as shown below.

\subsection{Limited BN range and value}

The in-memory BN mapping is the same as the weight mapping, which will map a bias value to a wordline of memory cells. For example, to map 32 to our 64x64 IMC array, half of a wordline of memory cells will store '1' as '+1', and the other half will store '0' as '-1'. The input for the in-memory BN is set to 1. Thus, $bias = \sum_{i=0}^{63}W_i$. For example, assume that $bias = W_0 + W_1+ W_2 + W_3$. If all $W_i = 1$, bias = 4. If $W_0 = -1$ and other $W_i = 1$, bias = -1 + 1 + 1 +1 = 2, Thus, the BN bias is even only if the width of the memory array is even (as in our case). Similarly, the BN bias will be odd only if the width of the memory array is odd.  To fit such constraints for in-memory BN mapping, we tried four different mapping methods: add, absolute add, sub, and absolute sub, on the target model. The one with the lowest accuracy drop will be selected as the choice. 

\begin{figure}[!!htb]
  \centering
  \includegraphics[height=!,width=1.0\linewidth,keepaspectratio=true]
  {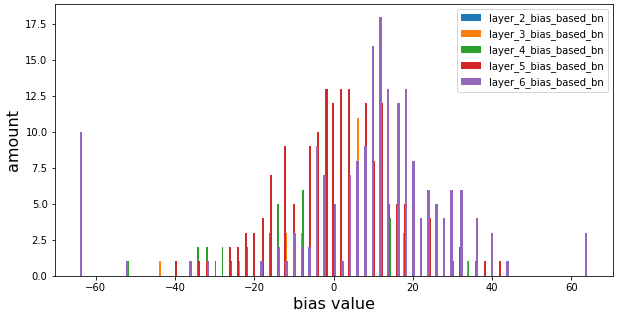}
  \caption[BN bias distribution of each layer.]{{ BN bias distribution of our KWS model. Best viewed in color.}}
  \label{fig:BN_distribution}
\end{figure}

Furthermore, the BN bias value will be limited to [-64, 64] due to the crossbar size of our IMC macro. To solve this problem, we first analyze the distribution of the BN bias, as shown in Fig.~\ref{fig:BN_distribution}. In this case, most of the BN bias does not exceed the limitation of [-64, 64], and thus the limited BN range has almost no impact on the accuracy of the model.

\subsection{MAV offset and SA sensing variation}

The convolution result of the IMC macro is not as ideal as in the software case due to the MAV offset and SA sensing variations. This will lead to a catastrophic failure of the model if we do not take any compensation measures. 
For the MAV offset, the MAV result is decided by the voltage difference between $AVG_p$ and $AVG_n$. This voltage difference shall be zero if the numbers of positive and negative values are the same in the ideal case. However, this difference (denoted as the MAV offset) is not zero due to the matching problem.
For SA sensing variation, the requirement of SA circuit input resolution is very high. Therefore, when the difference of two inputs is small, the variation may lead to the wrong comparison result. 

To solve above issues, we treat the MAV offset and SA variations as a random offset noise for inference, which is based on the Monte-Carlo simulation results with PVT variations. We applied this random noise to the model inference and compared the convolution results with the original ones to collect the statistics of their difference. A bias is then determined based on the statistics to restore the results as the original ones. This extra bias can be combined with the in-memory BN bias, since most of the BN bias values are within the limitation. After the compensation, we fine-tune the model for a few epochs, which could almost recover the accuracy drop due to these non-ideal effects.

\section{Proposed Architecture}
\subsection{Overall architecture}
\begin{figure}[!htb]
  \centering
  \includegraphics[height=!,width=1.0\linewidth,keepaspectratio=true]
  {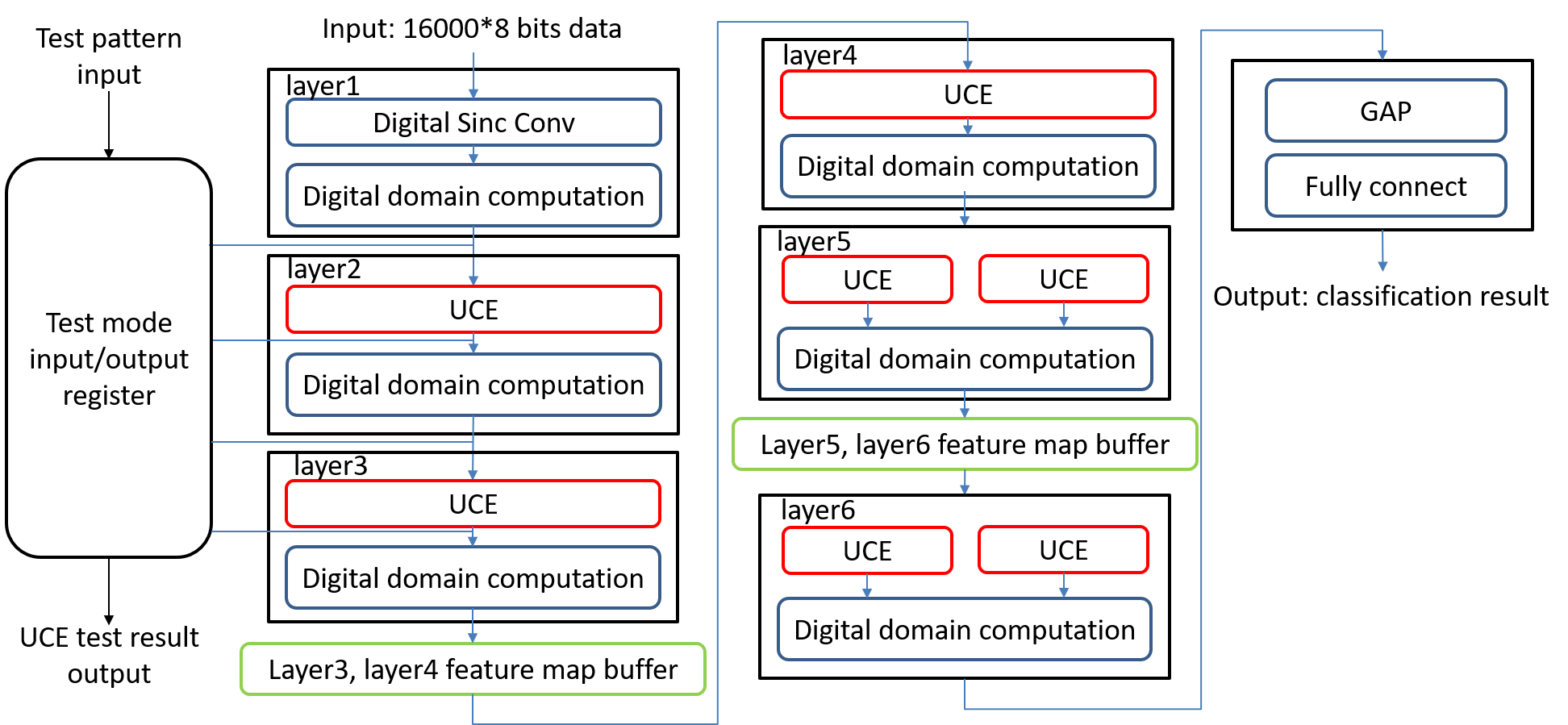}
  \caption[Overall hardware architecture for the inference part.]{{ Overall hardware architecture for the inference part. For better clarity, we only draw the test register connection to layer 2 and layer 3.}}
  \label{fig:hardware_architecture}
\end{figure}

\begin{figure}[!!htb]
  \centering
  \includegraphics[height=!,width=1.0\linewidth,keepaspectratio=true]
  {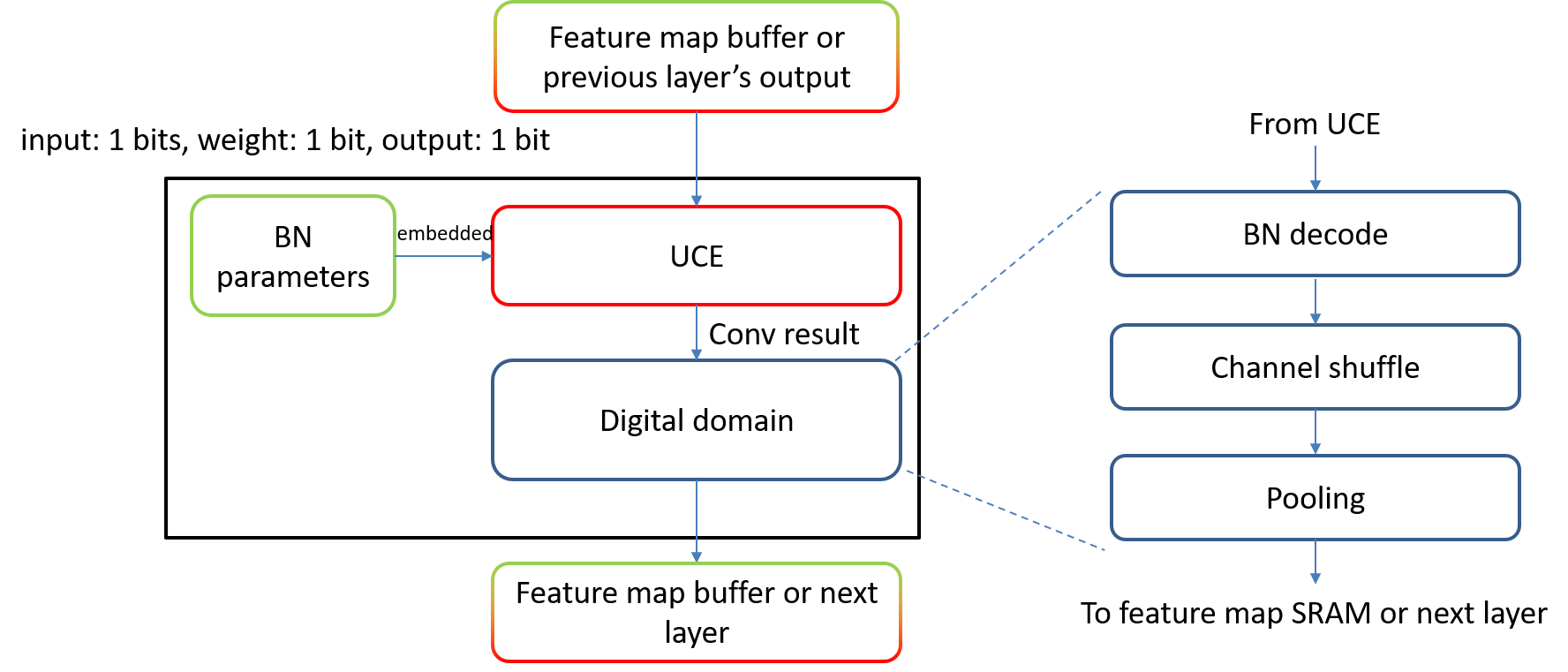}
  \caption[Layer block diagram for the inference part.]{{The block diagram for layers with IMC.}}
  \label{fig:layer_block_diagram}
\end{figure}

Fig.~\ref{fig:hardware_architecture} shows the overall architecture for our 6 layer KWS model. This design stores all weights in the IMC macros to avoid weight load/store overhead. For the model, we implement the first layer with digital circuits since the resolution of SA is too low to achieve the model requirement. Moreover, the final global average pooling (GAP) layer and fully connected layer are implemented by digital circuits as well for higher bit precision requirements. The other binarized group convolution layers are all implemented by the IMC macro as shown in Fig.~\ref{fig:layer_block_diagram}, which consists of an IMC macro for convolution and in-memory BN computation and digital domain computations such as BN decoder for correct sign operation, channel shuffle and pooing. The input data are from the previous layer or the buffer, and the output data is also output to the next layer or buffer. In addition, we have added a test mode to check the correctness and impact of non-ideal effects of each IMC macro, which can monitor the circuit variation of MAV and SA based on the input pattern and the result of each IMC macro. For our KWS task, the hardware utilization of each layer is (L1: 100, L2: 100, L3: 50, L4: 25, L5: 25, L6: 12.5) due to the pooling layer.  \par

\subsection{Sinc convolution circuits}

\begin{figure}[!!htb]
  \centering
  \includegraphics[height=!,width=1.0\linewidth,keepaspectratio=true]
  {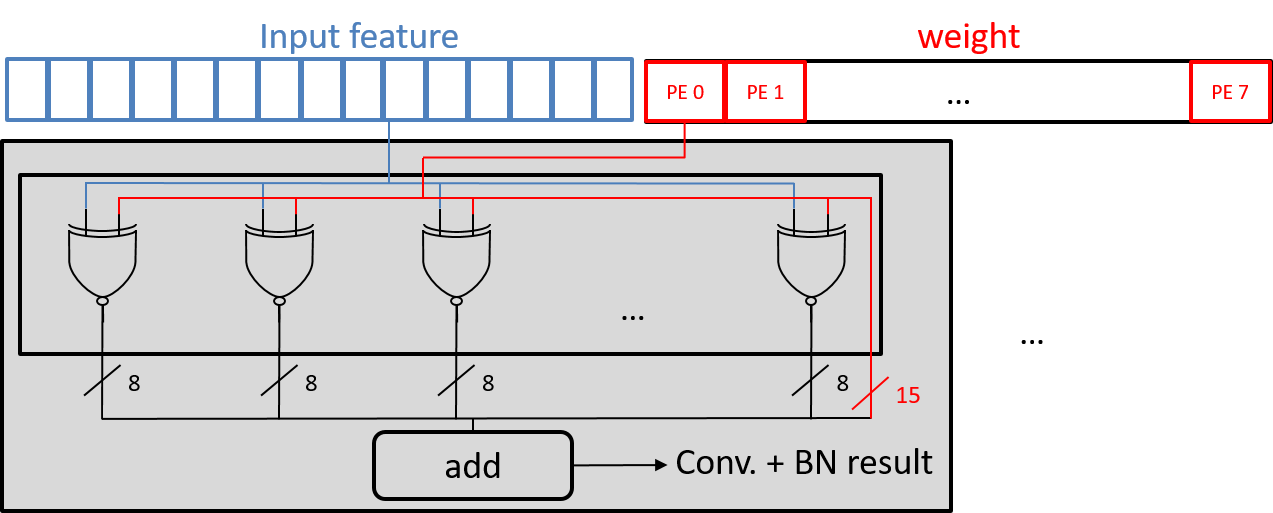}
  \caption[Digital Sinc convolution circuits.]{{ Digital Sinc convolution circuits.}}
  \label{fig:hardware_digital_sinc}
\end{figure}

Fig.~\ref{fig:hardware_digital_sinc} shows the block diagram for the sinc convolution layer, which consists of 8 PEs for 8 channel results in a single cycle to meet throughput requirements. Each PE computes 15(kernel size)$\times$8(input bitwidth) XNOR operations for binary multiplicaion, and accumulates them along with the BN bias as the channel output. The BN computation in this layer is also simplified as a bias value as the in-memory BN due to the binary output, which can be implemented with an adder instead of complex circuits.

\subsection{On-Chip Training circuits}

\begin{figure}[!htb]
  \centering
  \includegraphics[height=!,width=1.0\linewidth,keepaspectratio=true]
  {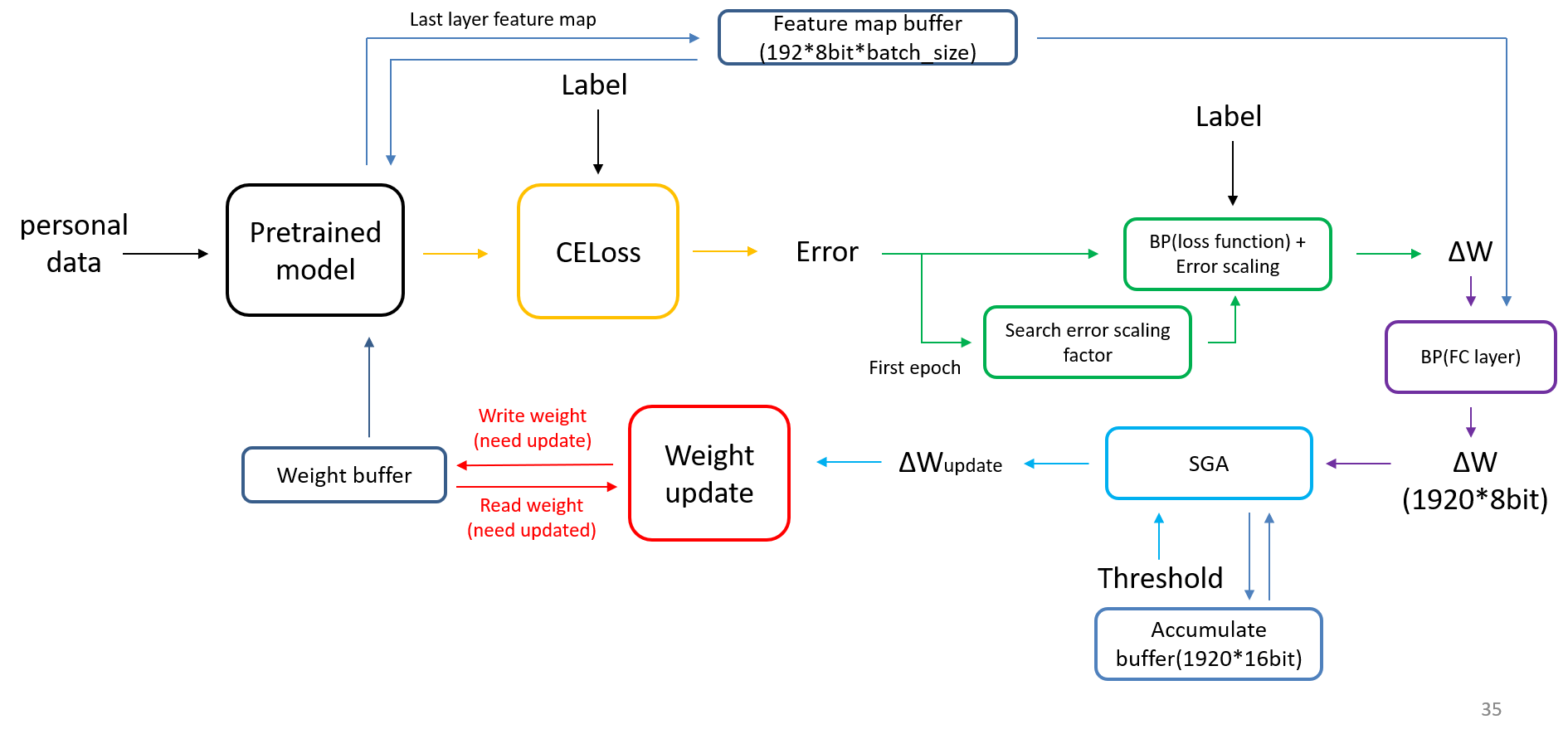}
  \caption[Overall flow for the training part.]{{ Overall flow for the training part.}}
  \label{fig:hardware_training_flow}
\end{figure}

\begin{figure}[!htb]
  \centering
  \includegraphics[height=!,width=1.0\linewidth,keepaspectratio=true]
  {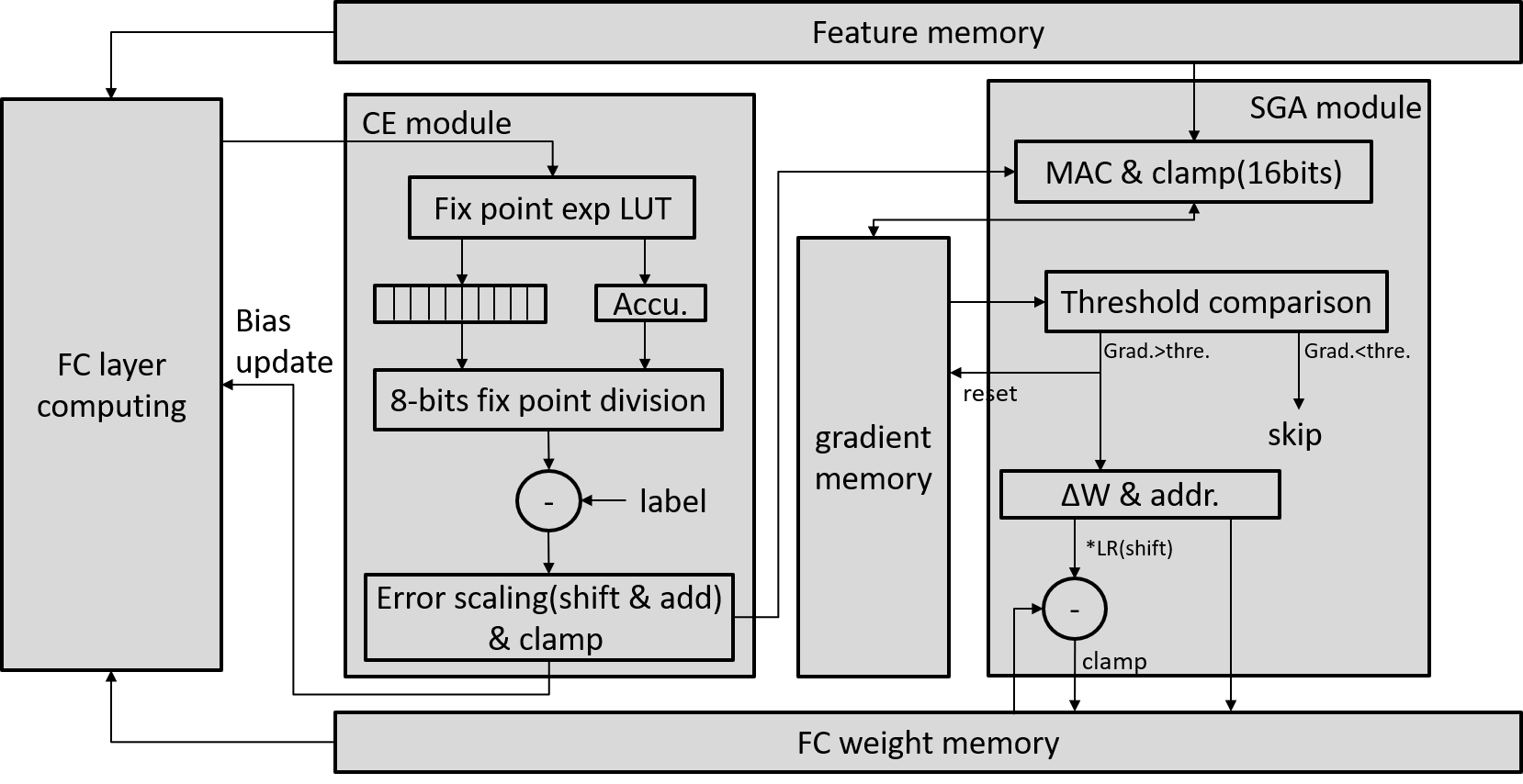}
  \caption[Block diagram for the training part.]{{ Block diagram for the training part.}}
  \label{fig:hardware_training_block_diagram}
\end{figure}

Fig.~\ref{fig:hardware_training_flow} shows the proposed on-chip training flow for model customization. Fig.~\ref{fig:hardware_training_block_diagram} shows the hardware block diagram of the training circuits.
First, the input feature for the last layer will be stored in an SRAM buffer for data reuse during the training process. Then the feature is read from the buffer for the last-layer inference to get the classification result. Based on the result, we calculate the error of the cross-entropy loss and its derivative for backpropagation. To avoid exponential computation in digital circuits, we replace it with a look-up table since the fully connected layer output are all low-precision fixed-point values. The look-up table can easily cover all situations with a small size register file. Furthermore, the division during the error calculation is fixed to 8 bits. \par

Once we get the backward error, we can apply the gradient scaling to avoid the zero-gradient situation, as mentioned before. The scaling factor here is treated as a hyperparameter which considers the distribution and the batch size to find a hardware-friendly scaling factor. In our work, since the scaling factor searched by the software simulation is 128, and the batch size is 90, the ideal scaling factor for hardware should be 1.42. The reason of the different scaling factors is that the error in Pytorch will be calculated in parallel and averaged in batch, but it is calculated sample by sample in hardware. Thus, the ideal hardware scaling factor should be \textit{the scaling factor of software/batch size}. This scaling factor is further simplified to 1.375 to replace multiplication with shift and add. \par

The error will be multiplied with the input to obtain the weight gradient and then stored in the gradient SRAM until the gradient computation for all data. These gradient values will be checked to see whether the value is higher than the threshold. Only sufficiently large gradients will be used to update the weights.

\section{Experiments and results}
\subsection{Results of Model Customization}
\subsubsection{Original dataset}
In our experiment, we use ten keywords in the Google speech command dataset (GSCD). They are: yes, no, up, down, left, right, stop, go, on, off. Each utterance is an audio file of nearly one second with a total of 18,947 training utterance and 4,735 test utterance.

\subsubsection{Personal dataset}
The personal data set is collected from three people with 607 utterance, where each utterance is also nearly one second. We use three utterance from each person as the training set for each keyword, which means that the customization training set will have 3 utterance x10 keywords x3 people = 90 training utterance. The rest are used as test. In our experiments, we only show customization for three people at the same time to mimic the real application scenario in a family.

\subsubsection{Settings}
For the original model, we randomly add Gaussian noise with values within 0.001 to 0.015 and randomly shift the audio by -0.5 to 0.5 seconds for data augmentation,  and train the model for 500 epochs with Adam optimizer. The initial learning rate is set to 0.01 and gradually decreased to the minimum value 1e-9 during training. \par

For model customization, we fine-tune the model using the SGD optimizer for 1000 epochs. The initial learning rate is set at 1/16, which is decreased every 10 epochs by a factor of 0.5 to the minimum value of 1/128. It should be noted that the learning rate cannot be set too low. Otherwise, the gradient value will be too small to update the parameters after multiplied by the learning rate. The quantization format for fine-tuning the classifier layer is listed below:
\begin{itemize}
\item weight: 1 sign bit, 7 decimal bits
\item activation: 1 sign bit, 3 integer bits, 4 decimal bits
\item gradient: 1 sign bit, 7 decimal bits
\item error: 1 sign bit, 7 decimal bits
\end{itemize}

\subsubsection{Performance of the original model}
Table~\ref{table:KS_result_comparison} shows the accuracy result of the original model on GSCD and related comparisons. Our compressed model uses a 7x smaller model size with more than 90\% accuracy due to the binary neural network compared to other state-of-the-art work. 

\begin{table}[!htb]
\centering
\caption[KWS accuracy(the ideal model).]{KWS accuracy(the ideal model).}
\label{table:KS_result_comparison}
\begin{tabular}{@{}cccc@{}}
\toprule
Model                                       & Accuracy             & Parameters           & Model Size (bits)     \\ \midrule
DS-CNN-S\cite{zhang2017hello}               & 94.1\%               & 39K                  & 1.2M                 \\
TC-ResNet8\cite{choi2019temporal}           & 96.1\%               & 66K                  & 2.1M                 \\
SincConv+GDSConv\cite{mittermaier2020small} & 96.4\%               & 62K                  & 2M                   \\ \midrule
Ours                                        & 90.83\%              & 125K                 & 171K                 \\ \bottomrule
\end{tabular}
\end{table}

Table~\ref{table:KS_result_comparison_hardware} shows the accuracy result of the original model in GSCD considering the hardware constraints mentioned previously. In which the FC quantized means that the weights in the fully connected layer is quantized to 8 bits, BN constraints include the influence of the limited range and limited value, and MAV offset and SA sensing variation are the non-ideal effects caused by the IMC macro. For the simulation with MAV offset and SA sensing variation, the results in the table are the averages of the five random seed simulations to cover the randomness. The result has shown that non-ideal effects will cause the model failure. However, accuracy can be restored after the proposed bias compensation and fine-tuning.

\begin{table*}[]
\centering
\caption[KWS accuracy(with hardware constraints).]{KWS accuracy(hardware constraints).}
\label{table:KS_result_comparison_hardware}
\begin{tabular}{@{}ccccccc@{}}
\toprule
Ideal                                 & \checkmark           &                      &               &            &              &                  \\
FC quantization                          &                      & \checkmark           & \checkmark    & \checkmark & \checkmark   & \checkmark       \\
BN constraints                        &                      &                      & \checkmark    & \checkmark & \checkmark   & \checkmark       \\
MAV offset + SA variation             &                      &                      &               & \checkmark & \checkmark   & \checkmark       \\
Bias compensation                     &                      &                      &               &            & \checkmark   & \checkmark       \\
Fine-tuning                           &                      &                      &               &            &              & \checkmark       \\ \midrule
Accuracy                              & 90.83\%              & 90.39\%              & 89.04\%       & 51.08\%    & 88.84\%      & \textbf{89.76\%} \\ \bottomrule
\end{tabular}
\end{table*}

\subsubsection{Customization}
Table~\ref{table:personalization_result} shows the customization result. In this case, the baseline (full-precision) is fine-tuned with full-precision on GPU as a reference. As shown in the table, naively fine-tuning on the quantized hardware will significantly degrade the performance. With the proposed method, we can achieve a much better performance close to the full precision one. 
In the case of training with RGP, the accuracy is higher than in the full precision baseline, which seems to be unreasonable. This could be the error caused by the insufficient amount of our test data, since the loss of the full precision baseline is still lower than the RGP one. 

Among the methods for on-chip customization, error scaling brings the largest gain, since it avoids training being early stopped. The accumulation of small gradients also improves the accuracy, which means that small gradients in the training process are important to improve the convergence of the model. Furthermore, for $\lambda$ in the random gradient prediction, our experiment shows that the value within a reasonable range (larger than 4) will not affect the result. Our method can make training on the fixed point hardware to get results comparable to that of the ideal model fine-tuning on GPU.

\begin{table}[!htb]
\centering
\caption[Model personalization result.]{Customization result with our methods.}
\label{table:personalization_result}
\begin{tabular}{@{}cccccc@{}}
\toprule
Baseline(FP) & \checkmark                &         &         &                  &                  \\
Quantized                &                  & \checkmark       & \checkmark       & \checkmark                & \checkmark                \\
Error scaling            &                  &         & \checkmark       & \checkmark                & \checkmark                \\
SGA                      &                  &         &         & \checkmark                & \checkmark                \\
RGP($\lambda$=8)         &                  &         &         &                  & \checkmark                \\ \midrule
Accuracy                 & \textbf{96.71\%} & 71.37\% & 86.46\% & \textbf{96.52\%} & \textbf{96.91\%} \\ \bottomrule
\end{tabular}
\end{table}

\subsection{Hardware Implementation Results}

\subsubsection{Results and comparison}
\begin{figure}[!htb]
  \centering
  \includegraphics[height=!,width=1.0\linewidth,keepaspectratio=true]
  {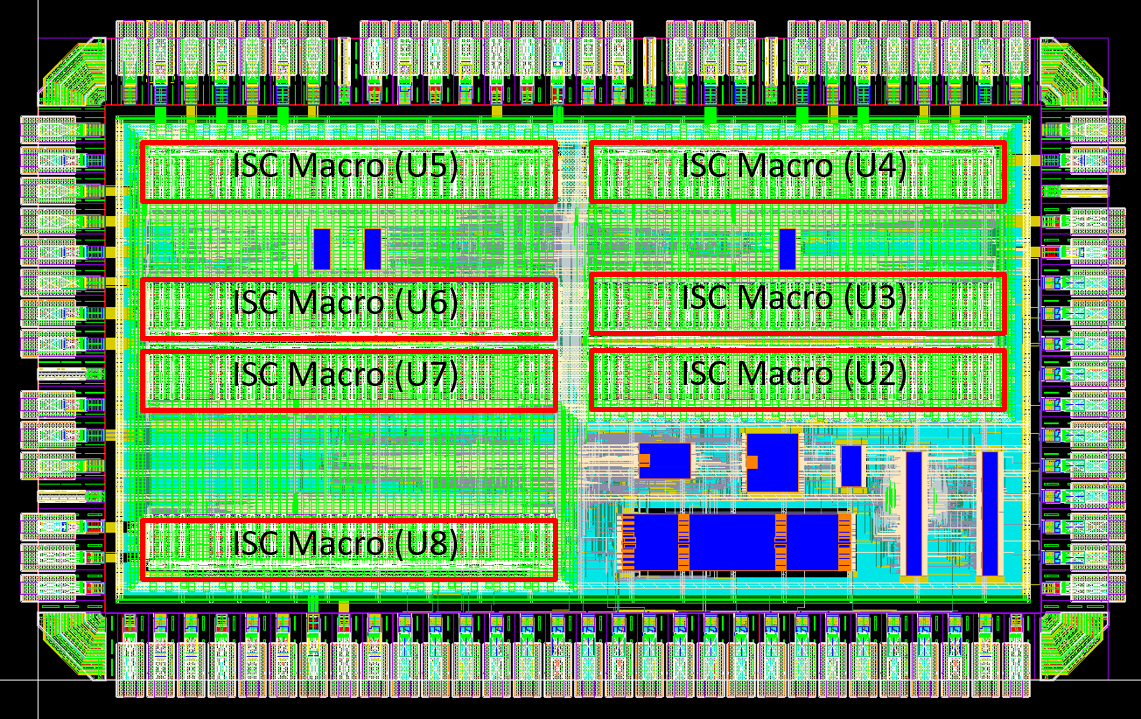}
  \caption[Layout diagram of the whole chip.]{{ Layout diagram of the whole chip.}}
  \label{fig:chip_layout_photo}
\end{figure}

\begin{figure}[!htb]
  \centering
  \includegraphics[height=!,width=0.5\linewidth,keepaspectratio=true]
  {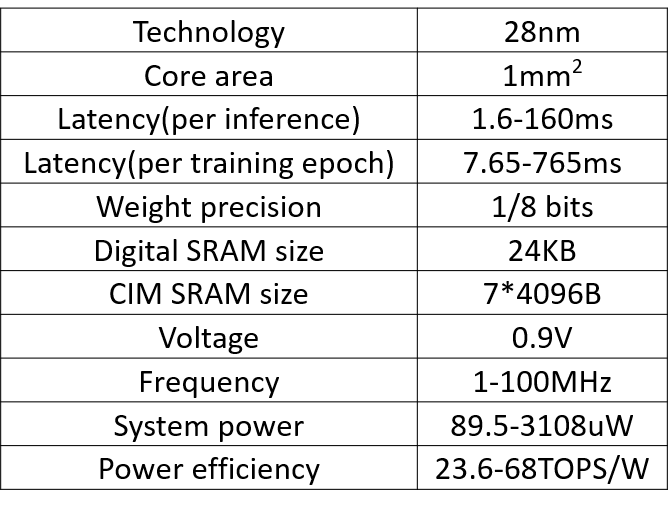}
  \caption[Chip summary.]{{ Chip summary.}}
  \label{fig:chip_summary}
\end{figure}

We have synthesized this design with Synopsys Design Compiler and performed the placement and routing with Cadence Encounder. All are in TSMC 28nm CMOS technology. Fig.~\ref{fig:chip_layout_photo} shows the layout of the chip, and Fig.~\ref{fig:chip_summary} shows the summary of this chip. This chip can work at different clock rates ranging from 1MHz to 100MHz at 0.9V. Timing and power consumption are evaluated at the TT corner. The power is analyzed by Synopsys PrimeTime PX based on the post-layout results and the gate-level simulation pattern of the KWS. This chip achieves 23.6-68 TOPS/W for the real model considering all on-chip power and real task inference time for 100MHz and 1MHz operating frequencies, respectively. The lowest power consumption is 89uW and 105uW in the inference and training phase on the 1 MHz clock.  

Table~\ref{fig:hardware_comparison} compares this work with other state-of-the-art works. Due to the difference of model and technology, the comparison is only possible to a limited extent.  Most of the other works used MFCC for feature extraction and VAD for voice activity detection\cite{zheng2019ultra,dbouk20200,guo20195,liu2019eera,liu2019ultra}. Since the VAD is for power savings, this could be integrated into this work as well. In this comparison,  this work consumes less power than other works with similar model architecture\cite{zheng2019ultra,liu2019eera,liu2019ultra} even though we process the entire raw data for the predicted results. \cite{dbouk20200,guo20195} use RNN as the model architecture, but \cite{dbouk20200} does not include feature extraction on the chip.

Another key point of this work is that we combine the IMC architecture for higher energy efficiency.  \cite{dbouk20200,guo20195} also uses in-memory computing and \cite{liu2019eera,liu2019ultra} uses approximate computation that is also an analog domain computation for lower power consumption. \cite{zheng2019ultra} is the full digital design counterpart for a similar application. Compared to these state-of-the-art works, this work has higher energy efficiency and on-chip training capability for customization.

\begin{table*}
\centering
\caption{Design comparison}
\label{fig:hardware_comparison}
\begin{tabular}{|l|l|l|l|l|l|l|} 
\hline
                                                                                                 & This work                                              & TCAS2019~\cite{zheng2019ultra}                                                      & ISSCC2020~\cite{dbouk20200}     & VLSI2019~\cite{guo20195}                                                                     & IEEE2019~\cite{liu2019eera} & IEEE2019~\cite{liu2019ultra}  \\ 
\hline
Technology                                                                                       & 28nm                            & 28nm                                                             & 65nm             & 65nm                                                                            & 28nm        & 22nm                                                                                                                                                                              \\ 
\hline
Algorithm                                                                                        & Sincnet + CNN                   & MFCC + CNN                                                       & MFCC + RNN       & MFCC + RNN                                                                      & MFCC + CNN  & MFCC + CNN                                                                                                                                                                        \\ 
\hline
Dataset(keyword number)                                                                          & GSCD(10)                        & \begin{tabular}[c]{@{}l@{}}TIMIT/TIDIGITS\\/HOME(1)\end{tabular} & GSCD(7)          & \begin{tabular}[c]{@{}l@{}}GSCD(10)/Hey snips(1) \\/Smart home(11)\end{tabular} & GSCD(10)    & GSCD(10)                                                                                                                                                                          \\ 
\hline
Accuracy(\%)                                                                                     & 89.76(96.52\textsuperscript{4}) & 95.3/98.6/96.0                                                   & 90.38            & 90.2/91.9/95.0                                                                  & 89.7        & 90.51                                                                                                                                                                             \\ 
\hline
Architecture                                                                                     & digital + IMC                   & digital                                                          & digital + IMC    & digital + IMC                                                                   & mixed mode  & mixed mode                                                                                                                                                                        \\ 
\hline
Weight/activation bits                                                                                     & 1/1                   & 1/1                                                          & mixed (4, 8)   & 1/1                                                                   & 1/(4, 8, 16)  & 7/8                                                                                                                                                                        \\ 
\hline
Core area(mm\textsuperscript{2})                                                                 & 1                               & 1.29                                                             & 4.13             & 6.2                                                                             & 0.94        & 0.75                                                                                                                                                                              \\ 
\hline
Normalized core area(mm\textsuperscript{2})\textsuperscript{1}                                   & 1                               & 1.29                                                             & 0.77             & 1.15                                                                            & 0.94        & 1.21                                                                                                                                                                              \\ 
\hline
SRAM buffer size (KB)                                                                                        & 24                            & -                                                                & 38             & 10                                                                            & -           & -                                                                                                                                                                                 \\ 
\hline
Frequency(MHz)                                                                                   & 1-100                           & 2.5-50                                                           & 1000             & 5-75                                                                            & 2.6         & 0.25                                                                                                                                                                              \\ 
\hline
Latency(ms)                                                                                      & 160-1.6                         & 0.5-10                                                           & 0.0399           & 0.127                                                                           & 20          & 20                                                                                                                                                                                \\ 
\hline
Power(uW)                                                                                        & 89-2833                         & 141                                                              & 11000            & 26000                                                                           & 77.8        & 52                                                                                                                                                                                \\ 
\hline
Normalized power(uW)\textsuperscript{2}                                                          & 89-2833                         & 328                                                              & 3838             & 9072                                                                            & 175         & 177                                                                                                                                                                               \\ 
\hline
Energy efficiency(TOPS/W)                                                                         & 23.6-68                         & 90                                                               & 0.91             & 11.7                                                                            & 137         & 46.8                                                                                                                                                                              \\ 
\hline
\begin{tabular}[c]{@{}l@{}}Normalized energy \\efficiency(TOPS/W)\textsuperscript{3}\end{tabular} & 23.6-68                         & 38.7                                                             & 2.6              & 33.5                                                                            & 60.9        & 13.7                                                                                                                                                                              \\ 
\hline
Remark                                                                                           & Customization                 & -                                                                & No preprocessing & -                                                                               & -           & -                                                                                                                                                                                 \\ 
\hline
\multicolumn{3}{l}{\begin{tabular}[c]{@{}l@{}}\textsuperscript{1}Normalized core area = Core area * 28\textsuperscript{2} / tech\textsuperscript{2}  \\ \textsuperscript{2}Normalized power = Power * (28 / tech) * (0.9\textsuperscript{2} / voltage\textsuperscript{2})\end{tabular}}   & \multicolumn{4}{l}{\begin{tabular}[c]{@{}l@{}}\textsuperscript{3}Normalized energy efficiency = energy efficiency / ((28 / tech) * (0.9\textsuperscript{2}~/ voltage\textsuperscript{2}))\\\textsuperscript{4}With model customization~ ~ ~ ~ ~ ~ ~ ~ ~ ~~\end{tabular}} 
\end{tabular}
\end{table*}

\subsubsection{Hardware analysis}
\begin{figure}[!htb]
  \centering
  \includegraphics[height=!,width=1.0\linewidth,keepaspectratio=true]
  {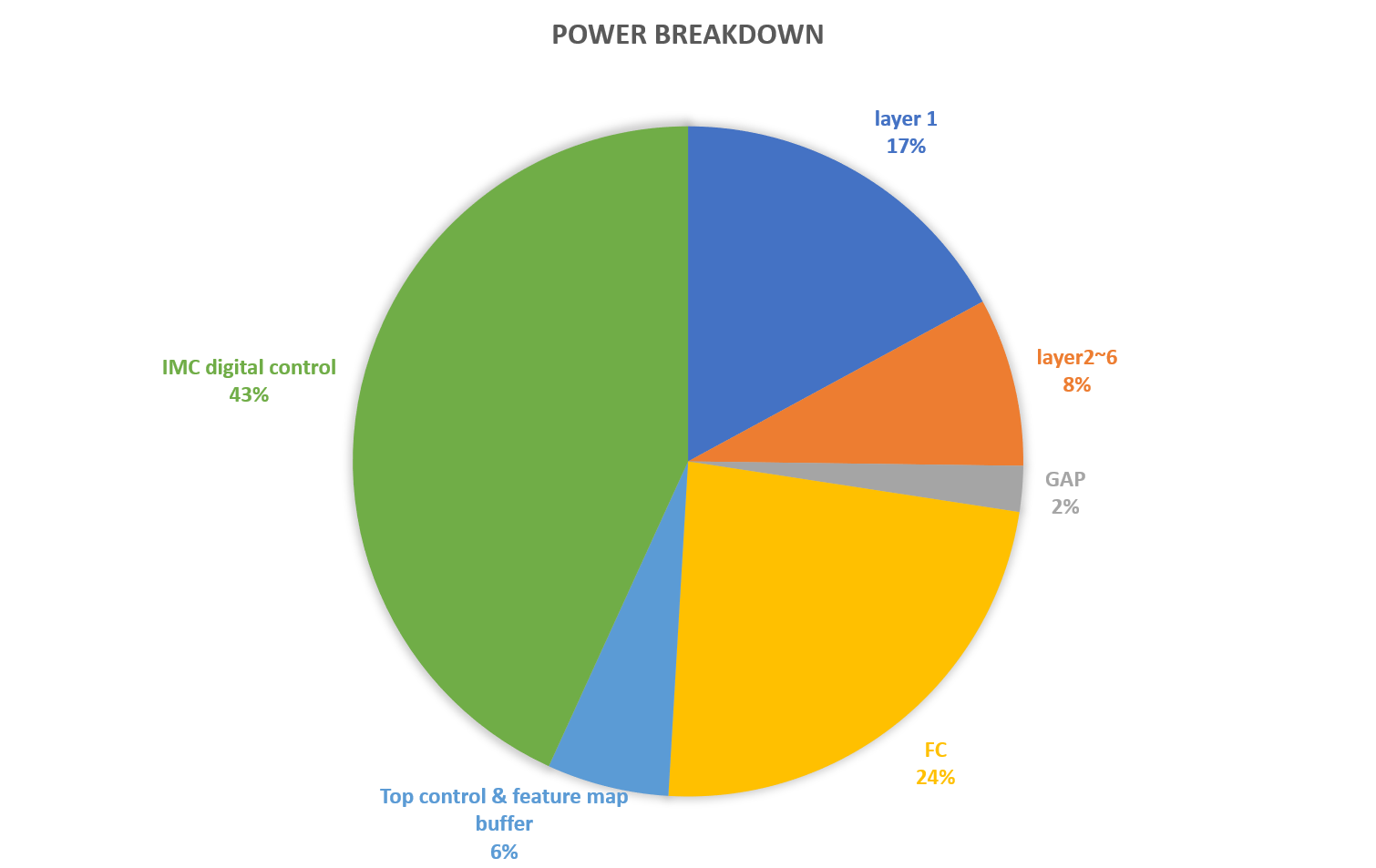}
  \caption[Power breakdown of each layer.]{{ Power breakdown of each layer.}}
  \label{fig:power_breakdown_layer}
\end{figure}

\begin{figure}[!htb]
  \centering
  \includegraphics[height=!,width=1.0\linewidth,keepaspectratio=true]
  {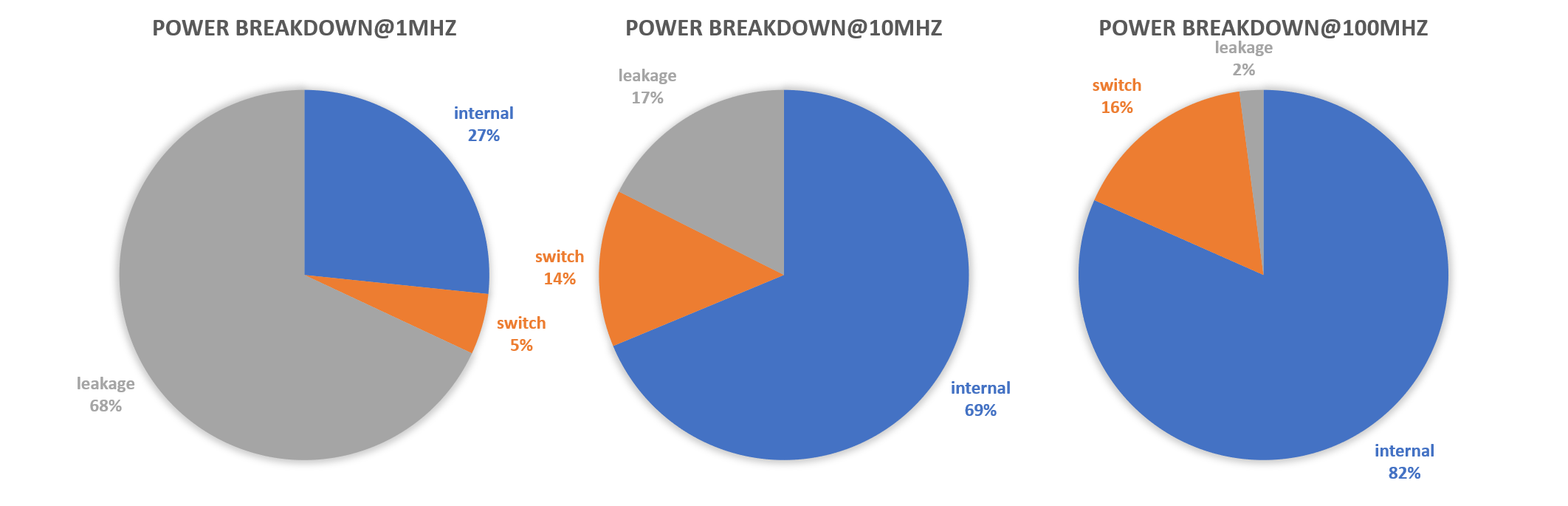}
  \caption[Power breakdown for different clock rates.]{{ Power breakdown of different clock rates.}}
  \label{fig:power_breakdown_clock}
\end{figure}

Fig.~\ref{fig:power_breakdown_layer} shows the power breakdown for model inference on the 1 MHz clock. In this case, most of the power is consumed by the fully connected layer and the IMC controller, since the fully connected part includes a large SRAM buffer and high-precision computation, and the IMC controller is implemented by many Flip-Flops. Therefore, the relative power consumption is higher. Furthermore, for higher throughput, the first layer needs more hardware overhead for greater parallelism, which occupies 18\% of the power. At the same time, the computation of the analog part consumes only 3\% of the total power. The leakage power will dominate the power consumption when the clock rate is low, as shown in Fig.~\ref{fig:power_breakdown_clock}. \par

\begin{figure}[!htb]
  \centering
  \includegraphics[height=!,width=1.0\linewidth,keepaspectratio=true]
  {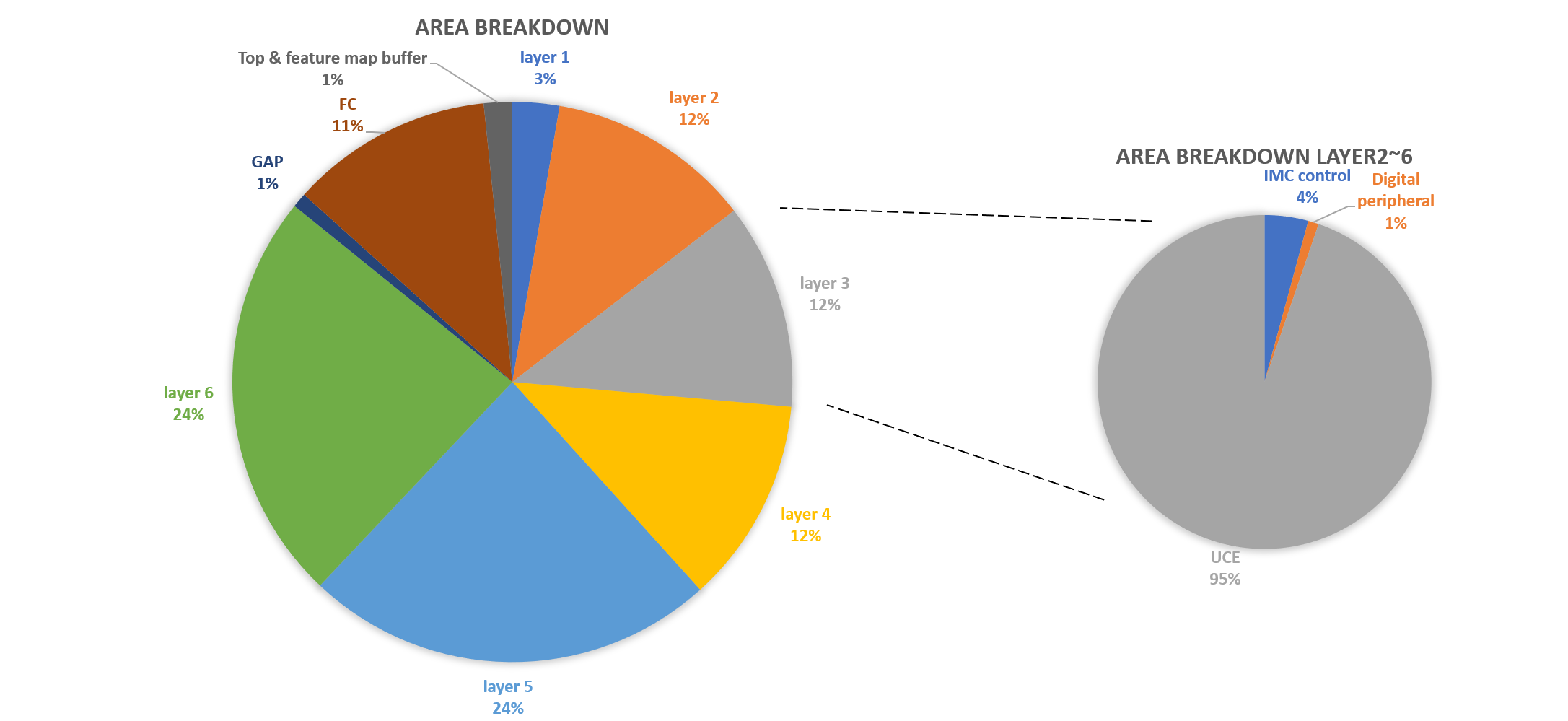}
  \caption[Area breakdown of each layer.]{{ Area breakdown of each layer.}}
  \label{fig:area_breakdown_layer}
\end{figure}

\begin{figure}[!htb]
  \centering
  \includegraphics[height=!,width=1.0\linewidth,keepaspectratio=true]
  {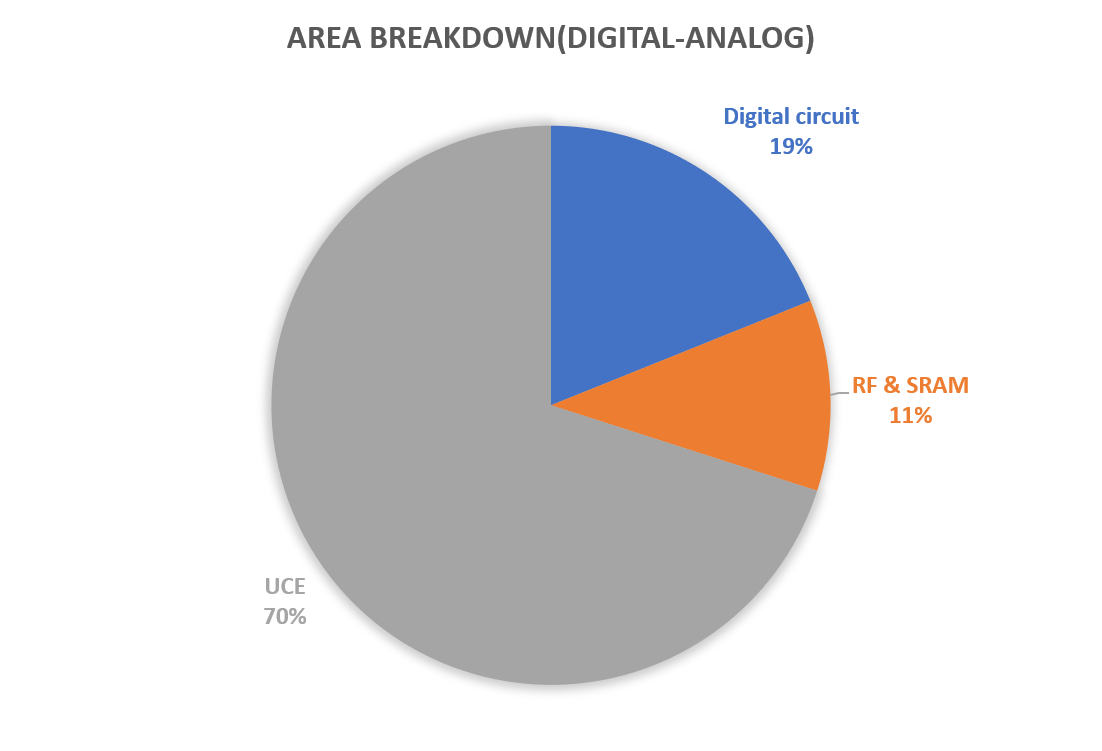}
  \caption[Area breakdown of digital and analog parts.]{{ Area breakdown of digital and analog parts.}}
  \label{fig:area_breakdown_ad}
\end{figure}

Fig.~\ref{fig:area_breakdown_layer} shows the area breakdown. In this case, the area is dominated by layer 5 and layer 6 since these two layers use 2 IMC macros, respectively, due to large numbers of parameters. For a single layer area, IMC macros dominate the area cost, while the IMC controller and digital peripheral circuits only cost about 5\% of the total area. Fig.~\ref{fig:area_breakdown_ad} shows the area breakdown for the analog and digital parts. The IMC macros cost about 70\% of the total area, the digital part circuits occypy 19\% of area, and the rest 11\% of area are the register file and the SRAM buffer. \par

In our implementation, the circuit for the training part only adds 5\% of the area in the overall design, around 9187 gate count, which means that our additional cost for the model customization function is relatively low. Within the training part circuits, the area is dominated by the large SRAM buffer for feature map storage. 

\section{Conclusion}
This paper presents a low power SRAM based IMC design with model customization for KWS, which is optimized from algorithm to hardware. For the algorithm, we propose an IMC aware model with fewer parameters to achieve over 90\% accuracy and solve the non-ideal effects of IMC macro with bias compensation and fine tuning. The model customization is designed to be executed on an 8-bit fixed-point quantized hardware. The limitation of the quantized hardware training is solved by scaling error, accumulating small gradients, and adding random gradients. The results show that the proposed approach can successfully restore accuracy and achieve a similar performance compared to fine-tuning with full precision. The hardware implementation uses hybrid digital/IMC computing to get better energy efficiency and fit model precision requirements, which has higher energy efficiency and also delivers on-chip model customization capability when compared to the state-of-the-art works.

\bibliographystyle{IEEEtran}

\bibliography{bib/IEEEabrv,bib/ieeeBSTcontrol, bib/thesis }

\begin{IEEEbiography}[{\includegraphics[width=1in,height=1.25in,clip,keepaspectratio]{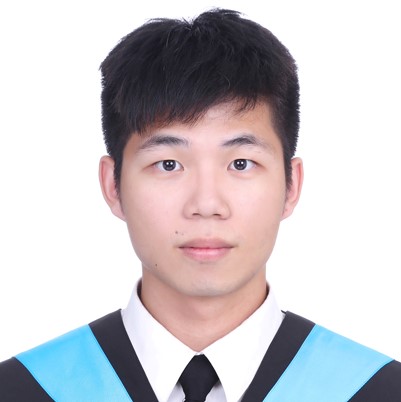}}]{Yu-Hsiang Chiang}
received the M.S. degree in electronics engineering from the National Yang Ming Chiao Tung University, Hsinchu, Taiwan, in 2021. He is currently working in the Novatek, Hsinchu, Taiwan. His research interest includes VLSI design and deep learning.

\end{IEEEbiography}

\begin{IEEEbiography}[{\includegraphics[width=1in,height=1.25in,clip,keepaspectratio]{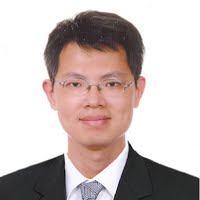}}]{Tian-Sheuan Chang}
	(S’93–M’06–SM’07)
	received the B.S., M.S., and Ph.D. degrees in electronic engineering from National Chiao-Tung University (NCTU), Hsinchu, Taiwan, in 1993, 1995, and 1999, respectively. 
	
	From 2000 to 2004, he was a Deputy Manager with Global Unichip Corporation, Hsinchu, Taiwan. In 2004, he joined the Department of Electronics Engineering, NCTU, where he is currently a Professor. In 2009, he was a visiting scholar in IMEC, Belgium. His current research interests include system-on-a-chip design, VLSI signal processing, and computer architecture.
	
	Dr. Chang has received the Excellent Young Electrical Engineer from Chinese Institute of Electrical Engineering in 2007, and the Outstanding Young Scholar from Taiwan IC Design Society in 2010. He has been actively involved in many international conferences as an organizing committee or technical program committee member.
\end{IEEEbiography}
\begin{IEEEbiography}[{\includegraphics[width=1in,height=1.25in,clip,keepaspectratio]{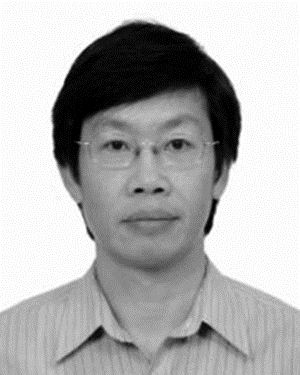}}]{Shyh-Jye Jou}
received the B. S. degree in electrical engineering from National Chen Kung University in 1982, and M.S. and Ph.D. degrees in electronics from National Chiao Tung University in 1984 and 1988, respectively. He joined Electrical Engineering Department of National Central University, Chung-Li, Taiwan,
from 1990 to 2004 and became a Professor in 1997. Since 2004, he has been Professor of Electronics Engineering Dept. of National Chiao Tung University and became the Chairman from 2006 to 2009. From August 2011 he becomes the Dean of Office of International Affairs, National Chiao Tung University. He was a visiting research Professor in the Coordinated Science Laboratory at University of Illinois, Urbana-Champaign during the 1993–1994 and 2010 academic years. In the summer of 2001, he was
a visiting research consultant in the Communication Circuits and Systems Research Laboratory of Agere Systems, USA. His research interests include
design and analysis of high-speed, low power mixed-signal integrated circuits, communications, and Bio-Electronics integrated circuits and systems.

Dr. Jou was the Guest Editor, IEEE JOURNAL OF SOLID STATE CIRCUITS, Nov. 2008. He served as the Conference Chair of IEEE International Symp. on VLSI Design, Automation and Test (VLSI-DAT) and International Workshop on Memory Technology, Design, and Testing. He also served as Technical Program Chair or Co-Chair in IEEE VLSI-DAT, International IEEE Asian Solid-State Circuit Conference, IEEE Biomedical Circuits and Systems, and other international conferences. He received Outstanding Engineering Professor Award, Chinese Institute of Engineers at 2011. He has published more than 100 IEEE journal and conference papers.
\end{IEEEbiography}
\end{document}